\documentclass[prb,showpacs,twocolumn,preprintnumbers,amsmath,amssymb,floatfix]{revtex4}
\usepackage[dvips]{graphicx}
\usepackage[latin1]{inputenc}
\begin{document}
\draft

\newcommand{\lxpc} {Li$_{x}$ZnPc }
\newcommand{\lpc} {Li$_{0.5}$MnPc }
\newcommand{\etal} {{\it et al.} }
\newcommand{\ie} {{\it i.e.} }
\newcommand{\ip}{${\cal A}^2$ }

\hyphenation{a-long}

\title{Magnetic properties of spin diluted iron pnictides from $\mu$SR and NMR in LaFe$_{1-x}$Ru$_x$AsO }

\author{Pietro Bonf\`a$^{1}$, Pietro Carretta$^{1}$,  Samuele Sanna$^{1}$, Gianrico Lamura$^{2}$, Giacomo Prando$^{1,3}$,
Alberto Martinelli$^{2}$, Andrea Palenzona$^{2}$, Matteo
Tropeano$^{2}$, Marina Putti$^{2}$ and Roberto De Renzi$^{4}$}

\address{$^{1}$ Department of Physics ``A. Volta'', University of Pavia-CNISM, I-27100 Pavia, Italy}
\address{$^{2}$ CNR-SPIN and Università di Genova, I-16146 Genova, Italy}
\address{$^{3}$ Department of Physics ``E.Amaldi'', University of Roma Tre-CNISM, I-00146 Roma, Italy}
\address{$^{4}$ Department of Physics, University of Parma-CNISM, I-43124 Parma, Italy}

\widetext

\begin{abstract}

The effect of isoelectronic substitutions on the microscopic
properties of LaFe$_{1-x}$Ru$_x$AsO, for $0\leq x\leq 0.8$, has
been investigated by means of $\mu$SR and $^{139}$La NMR. It was
found that Ru substitution causes a progressive reduction of the
N\'eel temperature ($T_N$) and of the magnetic order parameter
without leading to the onset of superconductivity. The temperature
dependence of $^{139}$La nuclear spin-lattice relaxation rate
$1/T_1$ can be suitably described within a two-band model. One
band giving rise to the spin density wave ground-state, while the
other one is characterized by weakly correlated electrons. Fe for
Ru substitution yields to a progressive decrease of the density of
states at the Fermi level close to the one derived from band
structure calculations. The reduction of $T_N$ with doping follows
the predictions of the $J_1-J_2$ model on a square lattice, which
appears to be an effective framework to describe the magnetic
properties of the spin density wave ground-state.

\end{abstract}

 \maketitle

\narrowtext

\section{Introduction}

The recent discovery of high temperature superconductivity nearby
the disruption of magnetic order in Fe-based
compounds\cite{Kamihara2008} has stimulated the scientific
community to further consider the role of magnetic excitations
\cite{Mazin2008} as a possible candidate for the pairing
mechanism. In order to address this point an appropriate
modellization of the microscopic magnetic properties of the Fe
based superconductors and of their parent compounds is necessary.
In the phase diagram of iron pnictides  superconductivity appears
to compete with a commensurate spin density wave (SDW) magnetic
order\cite{Luetkens2009a} characterized by a reduced magnetic
moment.\cite{Klauss2008} The magnitude of this moment is much
lower than that evaluated from band structure
calculations,\cite{Yildirim2008} possibly due either to the strong
electronic correlations, not appropriately taken into account in
those calculations, and/or from frustration effects.\cite{Si2008}

When the SDW phase is suppressed, by chemical substitution or by
applying high pressures, superconductivity is usually
recovered.\cite{Cruz2008,Luetkens2008,Okada2008} In fact, in
A(Fe$_{1-x}$Ru$_x$)$_2$As$_2$ (A=Sr, Ba) the substitution of Fe by
Ru suppresses the magnetic ordering for $x\rightarrow 0.3$ and
leads to bulk superconductivity for $0.2<
x<0.4$.\cite{Sharma2010,Schnelle2009,Rullier2010,Thaler2010} 
 On the other hand, it has been shown that in
PrFe$_{1-x}$Ru$_{x}$AsO,\cite{McGuire2009} in spite of an
analogous disruption of the SDW ordering for $x\simeq 0.6$, no
superconductivity is found up to the full Ru substitution. Indeed
band structure calculations for REFe$_{1-x}$Ru$_{x}$AsO show that
the electronic structure around the Fermi level is actually only
slightly affected by Ru substitution \cite{Tropeano2010} and that
only a minor charge doping takes place, even in the presence of
very large Ru contents.\cite{Tropeano2010} This means that in this
system the Fe/Ru substitution is effectively isoelectronic and
accordingly no relevant modification of the electronic
ground-state is observed. Furthermore it has been predicted
\cite{Tropeano2010} that the Ru atoms do not sustain any magnetic
moment suggesting that REFe$_{1-x}$Ru$_{x}$AsO should be
considered as a spin-diluted system.

In order to understand which is the effect of Ru substitution in
the 1111 family of iron pnictides we have performed $\mu$SR and
$^{139}$La NMR measurements in LaFe$_{1-x}$Ru$_{x}$AsO. A
progressive reduction of the N\'eel temperature ($T_N$) and of the
magnetic order parameter is observed with increasing $x$. Both
quantities eventually vanish for $x\rightarrow 0.6$ without
leading to the onset of superconductivity, at least up to $x=0.8$.
The temperature dependence of $^{139}$La nuclear spin-lattice
relaxation rate $1/T_1$ can be suitably described within a
two-band model, one giving rise to the SDW ground-state, while the
other one characterized by a Fermi-gas behaviour. Ru for Fe
substitution yields to a progressive decrease of the density of
states at the Fermi level, a trend which is quite consistent with
band structure calculations. The low temperature behaviour of
$1/T_1$ in the ordered phase and the reduction of $T_N$ with
doping can both be described within the $J_1-J_2$ model on a
square lattice and indicate that LaFe$_{1-x}$Ru$_x$AsO behaves as
a spin-diluted system with competing exchange interactions.  This
observation suggests that this model provides an effective
framework to suitably describe the role of  frustration in the
iron pnictides.\cite{Si2008,Johnst}

\section{Technical Aspects and Experimental Results}

The experiments were performed on polycrystalline
LaFe$_{1-x}$Ru$_x$AsO ($0\leq x\leq 0.8$) samples prepared as
described in Ref \onlinecite{Palecchi2011}. Structural
characterization was performed by X-ray powder diffraction at room
temperature and Rietveld refinement was carried out on selected
diffraction patterns. Microstructure was inspected by scanning
electron microscopy. Transport measurements show a decrease in the
resistivity $\rho$ with increasing Ru content and a shift to low
temperature of the characteristic anomaly in $d\rho/dT$ at
$T_N$.\cite{Palecchi2011} The magnetic susceptibility $\chi$
derived from SQUID magnetization measurements showed an analogous
behaviour of the peak in $d\chi/dT$ at $T_N$, shifting to lower
temperatures with increasing $x$. However, for $x\geq 0.4$ the
peak progressively smeared out and eventually for $x\rightarrow
0.6$ the susceptibility showed basically a Curie-Weiss behaviour.

\begin{figure}[h!]
\vspace{10cm} \includegraphics{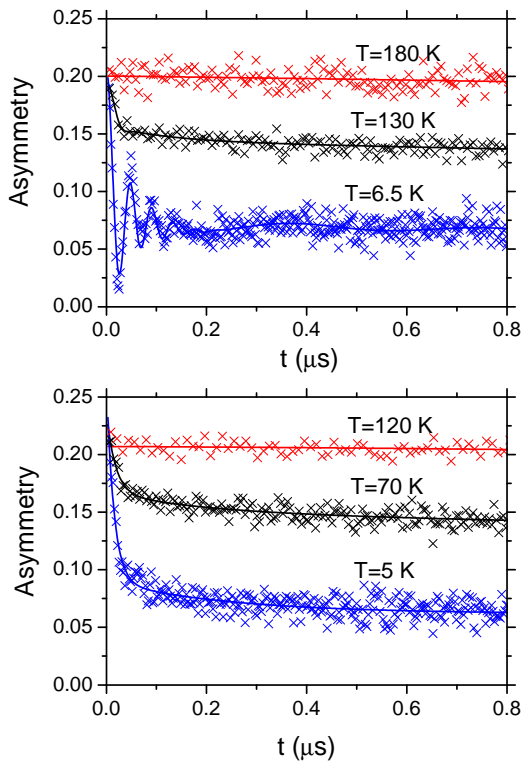} \caption{\label{Asym} Time evolution of
the muon asymmetry in LaFe$_{1-x}$Ru$_x$AsO for $x=0.1$ (top) and
for $x=0.4$ (bottom) at three selected temperatures. The solid
lines are the best fits according to Eq.\ref{ZFAsymm}. }
\end{figure}

Zero field (ZF) $\mu$SR experiments have been performed at PSI
with GPS spectrometer. For $T> T_N$, the muon asymmetry is
characterized by a decay which progressively increases with
decreasing temperature. Below $T_N$, at low Ru contents, well
defined oscillations are observed (Fig.\ref{Asym}), evidencing the
presence of a magnetic order. For $x\geq 0.3$ (Fig.\ref{Asym})
these oscillations are markedly damped due to the increase in the
local field distribution at the muon site.

\begin{figure}[h!]
\vspace{6cm} \includegraphics{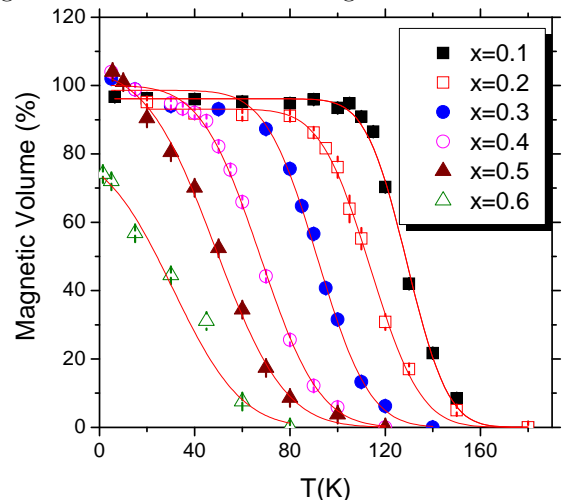} \caption{\label{VM} Temperature
dependence of the magnetic volume fraction in
LaFe$_{1-x}$Ru$_x$AsO. The solid lines are the best fits according
to the phenomenological expression $v_M(T)= 0.5 (1-erf(T-
T_N^{av}/\sqrt{2}\Delta))$.}
\end{figure}

Accordingly, for $x\leq 0.3$ the asymmetry could be fitted to the
sum of a fast and a slow oscillation plus a non-oscillating term,
namely
\begin{equation}
\label{ZFAsymm}
      A(t)= A_1 e^{-\lambda_1 t} f(\gamma_{\mu}B_1^{\mu}t)
      + A_2 e^{-\lambda_2 t} f(\gamma_{\mu}B_2^{\mu}t) +
      A_{\parallel} e^{-\lambda_{\parallel} t}
  \;\; ,
\end{equation}
where $\gamma_{\mu}$ is the muon gyromagnetic                                 
ratio, $B_{1,2}^{\mu}$ is the local field at the muon sites\cite{Takenaka2008,Maeter2009} $i=1$ or 2      
and $\lambda_{1,2}$ are the corresponding decay rates.  As can be seen      
from Fig.\ref{Asym} the amplitude of the fast oscillating               
component is significantly larger than the one of the slow
oscillating term. The fast oscillating signal could be reproduced
by $f(\gamma_{\mu}B_1^{\mu}t)= cos(\gamma_{\mu}B_1^{\mu}t+ \phi)$
for $x<0.3$. At higher Ru contents $f(\gamma_{\mu}B_1^{\mu}t)=
J_0(\gamma_{\mu}B_1^{\mu}t)$, with $J_0$ the zeroth order Bessel
function. For $T\ll T_N$, for $x< 0.6$, one finds that
$A_{\parallel}$ is 1/3 of the total asymmetry, as expected for
fully magnetically ordered powders.\cite{Amato1997} Then it is
possible to estimate the temperature dependence of the magnetic
volume fraction $v_M(T)= (3/2)(1- A_{\parallel}(T))$  from the
temperature dependence of $A_{\parallel}(T)$. The fraction
$v_M(T)$ is shown in Fig. \ref{VM} for different doping levels.
One notices a fast drop of $v_M$ for $T\rightarrow T_N$, which can
be empirically fitted with $v_M(T)= 0.5(1-erf(T-
T_N^{av}/\sqrt{2}\Delta_V))$, where $T_N^{av}$ represents an
average transition temperature. One observes a progressive
decrease of $T_N^{av}$ with $x$ together with some broadening,
likely due to inhomogeneity in the Ru distribution. Notice that
for $x\leq 0.5$ the temperature at which $v_M$ reaches about 100
\% is close to $T_N$ as determined from resistivity measurements,
as it is shown in details in the next section.

\begin{figure}[h!]
\vspace{6cm} \includegraphics{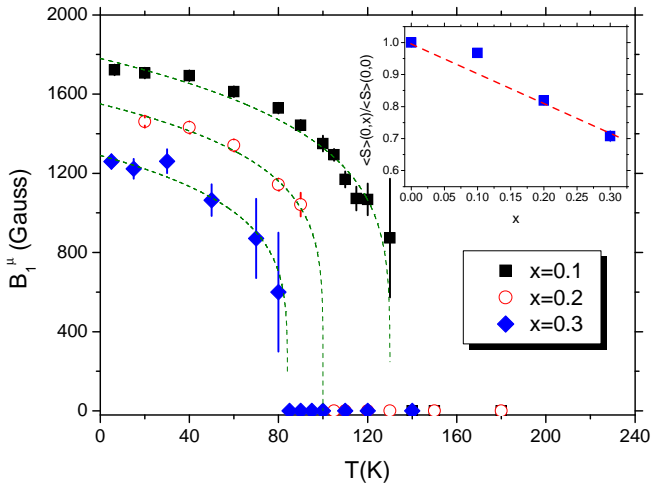} \caption{\label{Bmu} Temperature
dependence of the local field at the muon site 1 in
LaFe$_{1-x}$Ru$_x$AsO, for $0.1\leq x\leq 0.3$. The dashed lines
are guides to the eye. In the inset the $x-$dependence of the
$T\rightarrow 0$ order parameter $<S>(T\rightarrow 0
,x)/<S>(T\rightarrow0,0)= B_1^{\mu}(T\rightarrow
0,x)/B_1^{\mu}(T\rightarrow 0,0)$ is shown. The dashed line is a
guide to the eye.}
\end{figure}

The fit of the muon asymmetry for $x\leq 0.3$ allows one to derive
the temperature dependence of the modulus of the local field at
the muon $\mathbf{B}_i^{\mu}(T)= \mathcal{A}_i< \mathbf{S}
>$, with $\mathcal{A}_i$ the hyperfine coupling tensor and $<\mathbf{S}>$ the average Fe spin value, corresponding to
the order parameter. In Fig. \ref{Bmu} the temperature dependence
of $B_1^{\mu}(T)$, the internal field at the muon site close to
FeAs layers,\cite{Maeter2009} is reported for $x\leq 0.3$. One
notices a progressive decrease of the Fe magnetic moment with
increasing $x$.

\begin{figure}[h!]
\vspace{6cm} \includegraphics{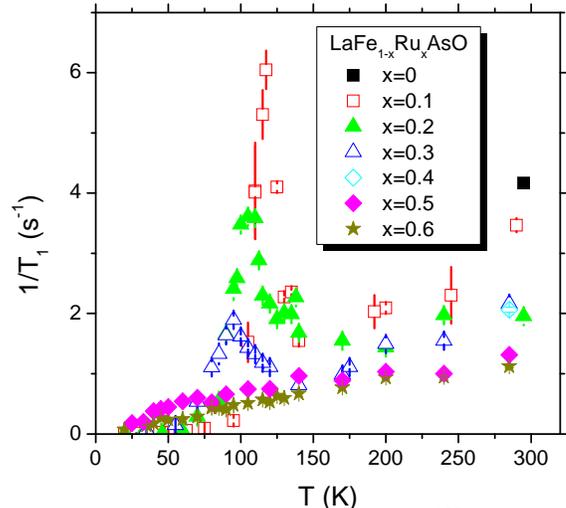} \caption{\label{T1vsT} Temperature
dependence of $^{139}$La $1/T_1$ in LaFe$_{1-x}$Ru$_x$AsO, for
$0.1\leq x\leq 0.6$, in a 7 Tesla magnetic field.}
\end{figure}

\begin{figure}[h!]
\vspace{7cm} \includegraphics{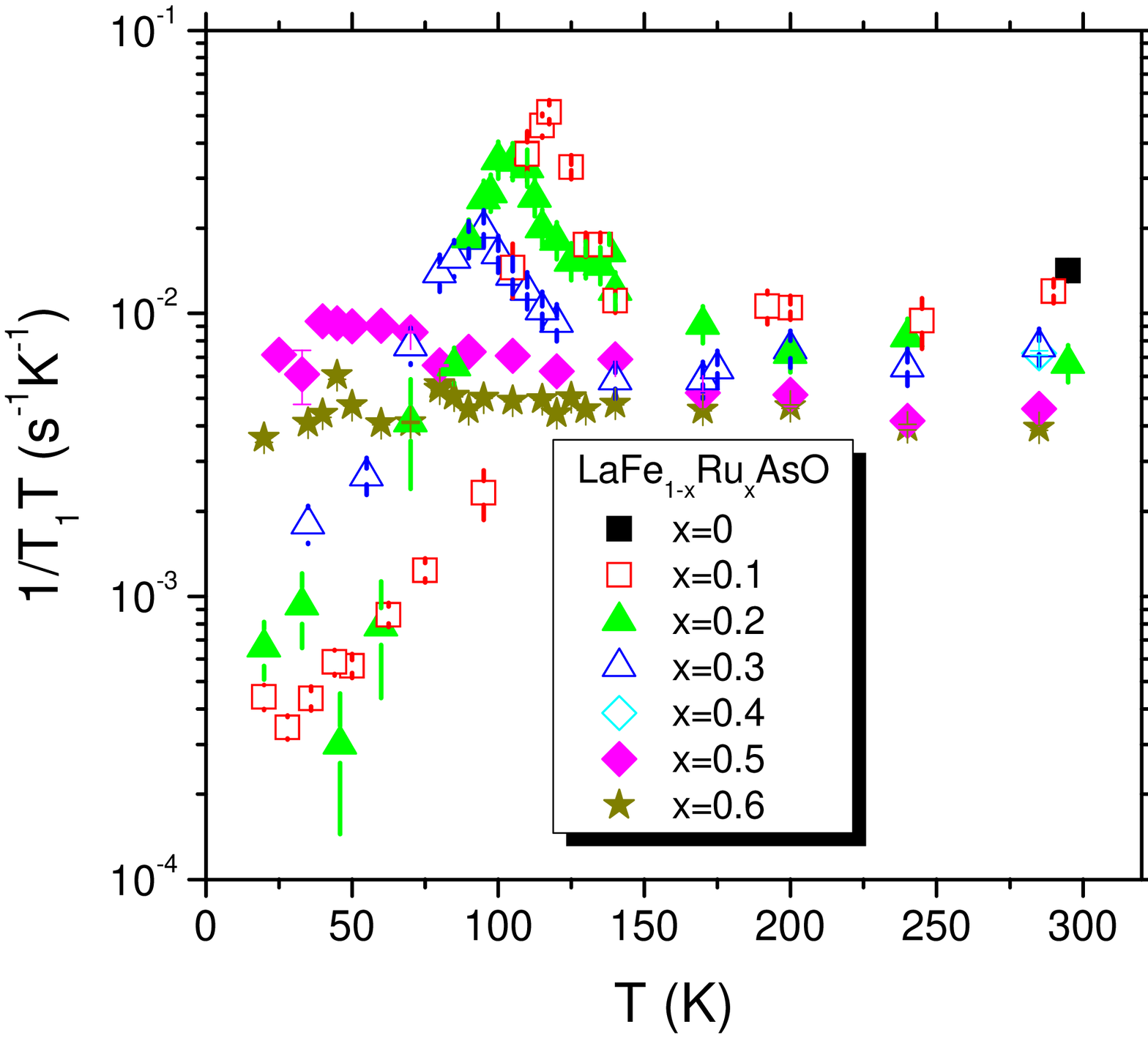} \caption{\label{T1TvsT} Temperature
dependence of $^{139}$La $1/T_1T$ in LaFe$_{1-x}$Ru$_x$AsO, for
$0.1\leq x\leq 0.6$, in a 7 Tesla magnetic field. }
\end{figure}

$^{139}$La NMR measurements have been carried out by using
standard radiofrequency pulse sequences. The spectra of the $x=0$
compound was in excellent agreement with the one previously
reported by Ishida \emph{et al.}\cite{Ishida2009} Upon doping one
observes that the peaks of the central line progressively smear
out and the spectrum gets narrower, suggesting a decrease in the
electric field gradient at $^{139}$La nuclei and possibly also a
change in the paramagnetic shift tensor. The explanation of this
phenomenology, however, goes beyond the aim of the present
manuscript. Nuclear spin-lattice relaxation rate $1/T_1$ was
measured on the central transition by using a saturation recovery
pulse sequence. The recovery of the nuclear magnetization does not
follow the trend expected  for a magnetic relaxation
mechanism\cite{Suter1998} by assuming a single $T_1$ but could
rather be fitted by assuming a distribution of $T_1$. Hereafter
$T_1$ will be defined as the time at which the recovery law decays
to $1/e$. The temperature dependence of $1/T_1$ in
LaFe$_{1-x}$Ru$_x$AsO is shown in Fig.\ref{T1vsT}. $1/T_1$ shows a
linear temperature dependence above $T_N$ and a drop below. The
peak at the transition temperature originates from the divergence
of critical fluctuations, progressively smeared out by
inhomogeneities for increasing $x$. Figure \ref{T1TvsT} displays
the temperature dependence of $1/T_1T$ on log-linear scales. It is
worth noticing that the relaxation rate does not vanish for
$T\rightarrow 0$, where it tends to a constant, $x-$dependent
value. The low temperature levelling of $1/T_1$ has already been
observed by Nakai et al. in Ref.\onlinecite{Nakai} for the parent
$x=0$ compound.

\section{Analysis and Discussion}

First we shall concentrate on the effect of Ru doping on the
sublattice magnetization and on the density of states at the Fermi
level in the light of band structure calculations, performed
within the density functional theory (DFT) using the local density
approximation (LDA) for the exchange and correlation functional.
The band structure was obtained with the {\sc Siesta}
code,\cite{soler2002siesta} which utilizes a linear combination of
atomic orbitals for valence electrons and separable norm
conserving pseudopotentials with partial corrections for atomic
cores. Crystal structures relaxations were performed with periodic
boundary conditions. The atomic positions as well as the cell
structure were allowed to be optimized in the paramagnetic state
by using a conjugate gradient algorithm. The real space
integration grid had a cut-off of 500 Ry and
 up to 12000 points were used for the Brillouin
zone sampling using the Monkhorst-Pack k-points sampling.
Stringent criteria were adopted for the electronic structure
convergence and  equilibrium geometry (residual forces lower than
$10^{-2}$ eV/\AA).

Our results are in close agreement with previous band structure
calculations reported in Ref. \onlinecite{Tropeano2010}. We find
that Fe magnetic moment decreases with Ru doping and eventually
vanishes for $x$ around 0.5. This trend is qualitatively
consistent with the one experimentally derived from the $x$
dependence of $B_1^{\mu}(T\rightarrow 0,x)$ (Fig.\ref{Bmu}).
Nevertheless, as it has already been pointed out in the
introduction, band structure calculations do not provide a
quantitatively correct estimate of the magnitude of Fe magnetic
moment, as well as of the $x$-dependence of the order parameter.
Namely, the initial slope $d<S(T\rightarrow 0,x)>/dx$ obtained
experimentally is much faster than the one derived from band
structure calculations.

\begin{figure}[h!]
\vspace{5cm} \includegraphics{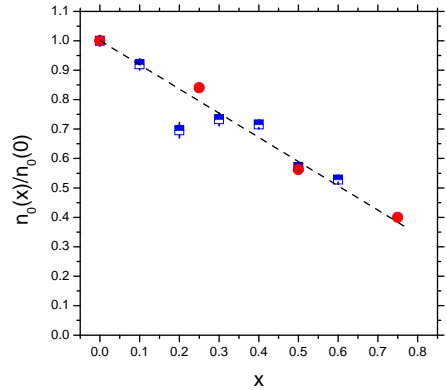} \caption{\label{DOS} The density of states
at  the Fermi level is reported as a function of Ru content in
LaFe$_{1-x}$Ru$_{x}$AsO, normalized to the $x=0$ value. The
squares represent the data derived from $^{139}$La $T_1$
measurements, while the circles are from band structure
calculations. The dashed line is a guide to the eye.}
\end{figure}

On the other hand, a better agreement with the experimental
findings is observed for the calculated density of states at the
Fermi level. This quantity can be derived experimentally from
$1/T_1$ measurements.\cite{slichter1990principles} In fact, above
$T_N$ $1/T_1$ follows the Korringa behaviour expected for a Fermi
liquid (Fig.\ref{T1vsT}), namely $1/T_1= C n_0^2 T$, with $n_0$
the density of states at the Fermi level and $C$ a constant
accounting for the hyperfine coupling between the electrons and
$^{139}$La nuclei. Then, by taking the value of the spin-lattice
relaxation rate around room temperature one can write that
$\sqrt{T_1(0)/T_1(x)}\simeq n_0(x)/n_0(0)$ and derive the
$x$-dependence of the density of states at the Fermi level. In
Fig.\ref{DOS} this ratio is compared to the results obtained from
\emph{ab initio} calculations in the paramagnetic state. One
observes that $n_0(x)$ decreases with increasing Ru content, in
excellent agreement with band structure calculations. This
decrease should be associated with the larger extension of Ru $d$
orbitals which leads to an enhanced delocalization of the
electrons.

At low temperature, in the SDW phase, LaFeAsO nuclear spin-lattice relaxation can be described as the sum of two
contributions, as suggested by Smerald \emph{et al.}\cite{Smerald2010}
\begin{equation}
\label{T1sum}
      \frac{1}{T_1}= \frac{1}{T_1^{FL}} + \frac{1}{T_1^{SW}}
  \;\; ,
\end{equation}
where $1/T_1^{FL}\propto T$ is a Fermi-gas like term arising from
weakly correlated electrons which accounts for the low-temperature
levelling of $1/T_1T$, while $1/T_1^{SW}$ is the one from a band
with strongly correlated electrons where the Fermi surface nesting
leads to the insurgence of the SDW phase.

At low temperature $1/T_1^{SW}$ reduction is determined by the gap
$\Delta$ in the spin excitations. In fact, following
Ref.\onlinecite{Smerald2010}, taking into account that the
experiments were performed on powders, one can write
\begin{eqnarray}
\label{T1SDW}
      \frac{1}{T_1^{SW}} \simeq \frac{4\mathcal{A}_h^2m_0^2\hbar
      V^2\gamma^2_N\Delta^3}{3\pi^3\chi_{\perp}^2v_s^6}\Phi\biggl[\frac{K_BT}{\Delta}\biggr]=\\ \nonumber
      = \frac{4}{3}\mathcal{A}_h^2\gamma^2_N m_0^2\hbar\alpha^2\Delta^3\Phi\biggl[\frac{K_BT}{\Delta}\biggr] \\
\mathrm{where}\,\,\, \Phi[x]= x^2 Li_1(e^{-1/x}) + x^3 Li_2(e^{-1/x}) \nonumber
\end{eqnarray}
with $Li_n(z)$ the $n^{th}$ polylogarithm of $z$. In
Eq.\ref{T1SDW} $\mathcal{A}_h$ is the hyperfine coupling between
the longitudinal fluctuations of the Fe moment and the nuclear
spin, $\gamma_N$ is $^{139}$La gyromagnetic ratio, $m_0$ the
amplitude of the fluctuating moment, while $\alpha=
V/\chi_{\perp}v_s^3\pi^{3/2}$, with $V$ the unit cell volume,
$v_s$ the average spin-wave velocity and $\chi_{\perp}$ the
transverse spin susceptibility. It is interesting to observe that
this approach applies also to the $J_1-J_2$ model on a square
lattice with localized spins.\cite{SmeraldPrivate} In fact,
although LaFe$_{1-x}$Ru$_{x}$AsO is not a localized spin system,
the $J_1-J_2$ model appears to be still applicable in some
effective version also to the iron pnictides. Further support to
this idea will be presented subsequently in the discussion of
LaFe$_{1-x}$Ru$_{x}$AsO phase diagram.

\begin{figure}[h!]
\vspace{10cm} \includegraphics{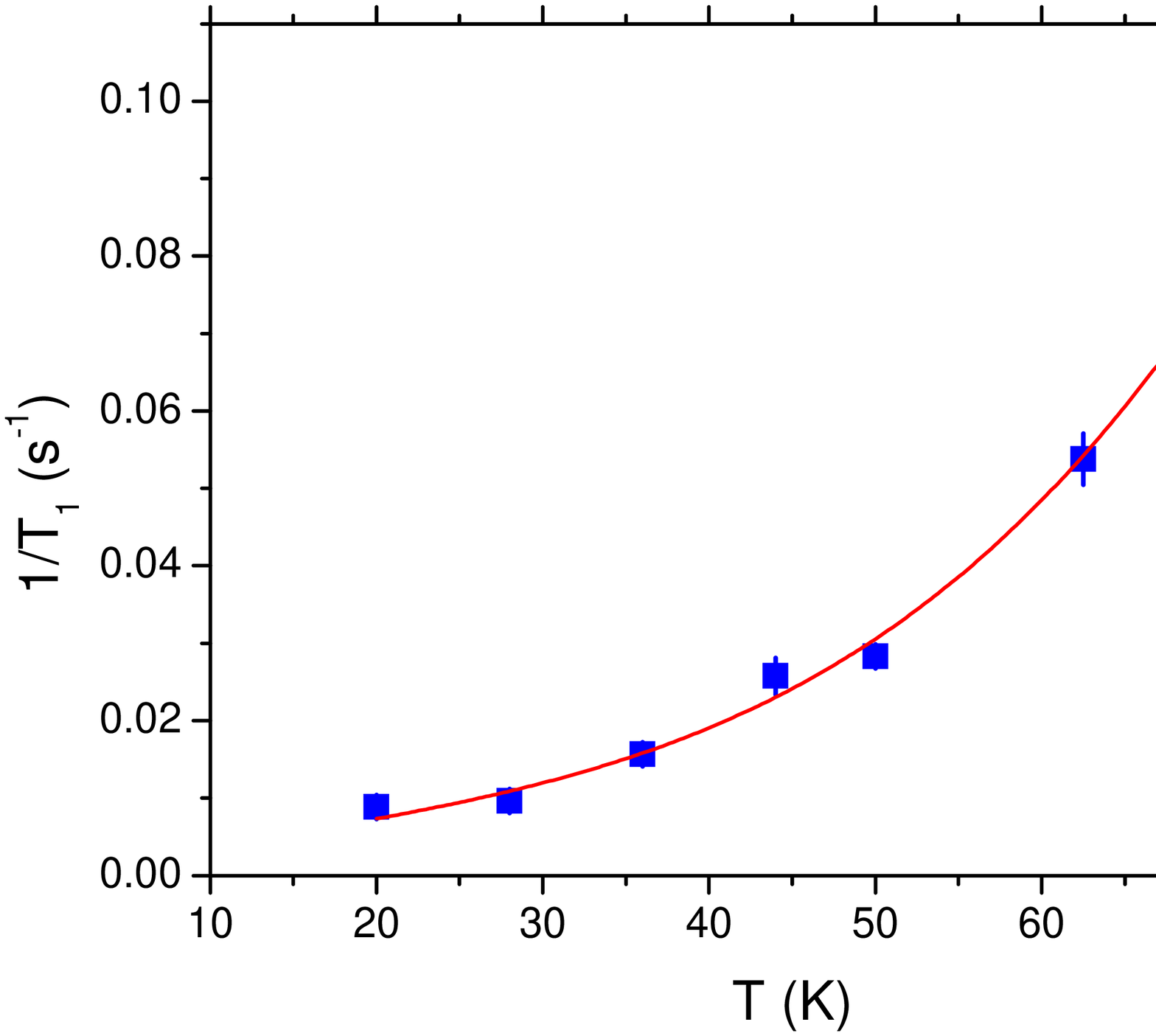} \includegraphics{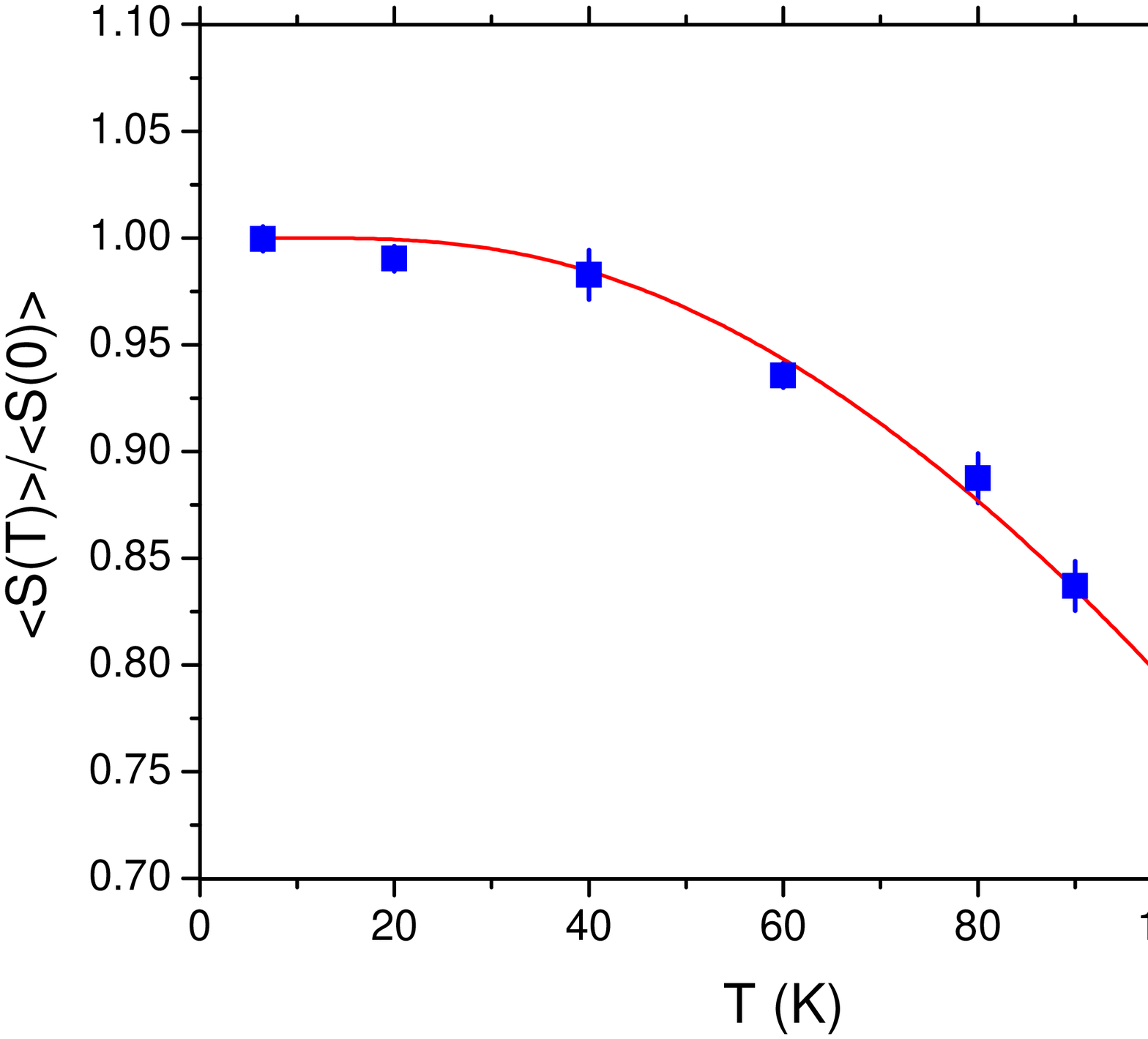} \caption{\label{Delta} (Top)
Temperature dependence of $^{139}$La $1/T_1$ in
LaFe$_{0.9}$Ru$_{0.1}$AsO, in a 7 Tesla magnetic field, for $T\ll
T_N$ . The solid line shows the best fit according to
Eq.\ref{T1SDW}.  (Bottom) Temperature dependence of the magnetic
order parameter derived from ZF-$\mu$SR measurements. The solid
line is the best fit according to Eq.\ref{MSDW}.}
\end{figure}

It is interesting to notice that many parameters appearing in Eq.\ref{T1SDW} determine also the reduction of the
sublattice magnetization. In fact one has that
\begin{equation}
\label{MSDW}
      \frac{<S>(0)-<S>(T)}{<S>(0)}= \alpha \sqrt{\Delta} \biggl(\frac{K_BT}{2}\biggr)^{3/2} e^{-\Delta/k_BT}
\end{equation}
Thus, both the temperature dependence of $<S>(T)$ and of $1/T_1$
for $T\ll T_N$ are determined by $\Delta$. By fitting ZF-$\mu$SR
and $^{139}$La $1/T_1$ curves for $x=0.1$ (Fig.\ref{Delta}) one
finds an accurate fit of both quantities for $\Delta= 8\pm 2$ meV,
a value which is close to the one found in other iron pnictides by
means of inelastic neutron
scattering.\cite{McQueeney2008,Zhao2008,Matan2009} By assuming in
Eq.\ref{T1SDW} $m_0\simeq 1 \mu_B$ one finds a quantitative
agreement with $^{139}$La $1/T_1$ data for $\mathcal{A}_h\simeq
1.8$ kG/$\mu_{\rm B}$. This value is reasonably close to that
estimated in the SDW phase from the hyperfine field  at the
$^{139}$La nucleus $B_{La}\simeq 2.4$ kG. Namely, by assuming the
$T=0$ value of the magnetic moment of Fe in the SDW
phase\cite{RDR} to be $m=0.6\,\mu_B$,  one has
$\mathcal{A}_h\simeq |{B_{La}}|/z<M(0)>\simeq 1$ kG/$\mu_{\rm B}$,
with $z=4$ the nearest neighbor Fe atoms.

\begin{figure}[t!]
\vspace{6.5cm} \includegraphics{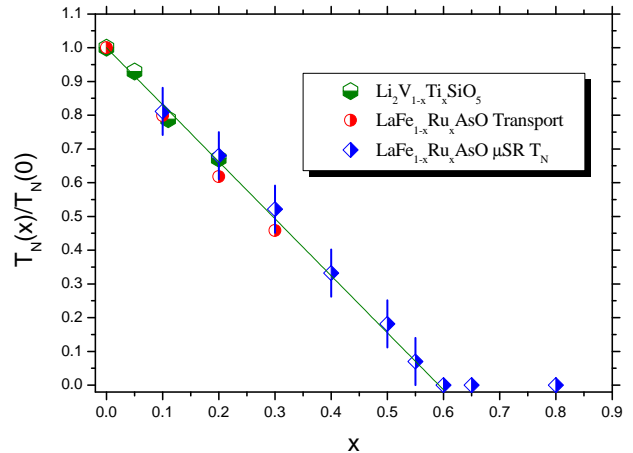} \caption{\label{PD} The doping dependence
of  N\'eel temperature in  LaFe$_{1-x}$Ru$_x$AsO, normalized to
its $x=0$ value is shown. Circles show the behaviour derived from
transport data, squares from ZF-$\mu$SR, while diamonds show the
corresponding behaviour for Li$_2$V$_{1-x}$Ti$_x$SiO$_5$.}
\end{figure}

Further insights on the applicability of the $J_1$-$J_2$ model to
LaFe$_{1-x}$Ru$_{x}$AsO comes from the analysis of the phase
diagram. As it is shown in Fig.\ref{PD} the N\'eel temperature
determined either from transport or from ZF-$\mu$SR by taking the
temperature at which $v_M\rightarrow 1$ (Fig. 2), decreases almost
linearly with $x$ and eventually vanishes around $x=0.6$. It may
be argued that $x\neq 0$ samples present a distribution of N\'eel
temperatures  and that the criteria chosen for identifying $T_N$
may vary with the determination technique. Still, if we choose,
for example, to evaluate the average $T_N^{av}$  from the flex in
the muon magnetic fraction $v_m(T)$, the slope $dT_N^{av}/dx$
turns out to be the same as that of Fig. \ref{PD}. This points out
that our main findings are not affected by some distribution in
the N\'eel temperatures.

Now, it is instructive to compare the slope $dT_{N}(x)/dx$ with
the one found in Li$_2$V$_{1-x}$Ti$_x$SiO$_5$, a prototype of the
$J_1-J_2$ model on a square lattice with spin dilution arising
from the substitution of V$^{4+}$ ($S=1/2$) with Ti$^{4+}$
($S=0$). One notices that the trend of $T_N(x)/T_N(0)$ is the same
in LaFe$_{1-x}$Ru$_{x}$AsO and Li$_2$V$_{1-x}$Ti$_x$SiO$_5$,
further supporting the idea that the $J_1-J_2$ model on a square
lattice\cite{Johnst} provides an effective framework to
appropriately describe the effects of the spin dilution induced by
Ru for Fe substitution.\cite{Papinutto2005} Moreover, it is
pointed out that in LaFe$_{1-x}$Ru$_{x}$AsO $T_N(x)$ vanishes for
$x$ around $0.6$ which corresponds to the percolation threshold
for the $J_1-J_2$ model on a square lattice.


\section{Conclusions}

In conclusion, we have shown that Ru for Fe substitution in
LaFe$_{1-x}$Ru$_{x}$AsO causes a progressive reduction of the
N\'eel temperature ($T_N$) and of the magnetic order parameter
without leading to the onset of superconductivity. The analysis of
$^{139}$La nuclear spin-lattice relaxation rate $1/T_1$ indicates
that this system can be described within a two-band model, one of
them giving rise to the spin density wave (SDW) ground-state. Fe
for Ru substitution yields to a progressive decrease of the
density of states at the Fermi level in quantitative agreement
with band structure calculations. The behaviour of $1/T_1$ in the
SDW phase and the reduction of $T_N$ with Ru substitution can both
be described within the $J_1-J_2$ model on a square lattice which
suggests that LaFe$_{1-x}$Ru$_x$AsO behaves as a spin-diluted
system with competing exchange interactions, pointing out the
relevant role of frustration in the parent compounds of iron based
superconductors.


The  assistance by Alex Amato during the $\mu$SR measurements at
PSI and the access support by EU contract RII3-CT-2003-505925
(NMI3) are gratefully acknowledged.



\end{document}